\documentclass{WileyMSP-template}
\usepackage{mhchem}
\usepackage{siunitx}
\usepackage{subcaption}
\usepackage{float}
\usepackage{booktabs}
\usepackage{mathtools}
\usepackage{multirow}
\DeclarePairedDelimiter\abs{\lvert}{\rvert}
\usepackage{comment}
\newcommand{\vms}{$\Delta V_{2ms}$}
\newcommand{\vct}{$\Delta V_{ct}$}
\newcommand{\vd}{$\Delta V_{d}$}
\newcommand{\macm}{mAhcm$^{-3}$}
\newcommand{\lix}{Li$_x$}
\newcommand{\mahg}{mAhg$^{-1}$}
\newcommand{\gcm}{gcm$^{-3}$}
\newcommand{\dqdv}{d$Q$d$V^{-1}$}
\newcommand{\mss}{m$^2$s$^{-1}$}
\newcommand{\refli}{Li/Li$^+$}
\newcommand{\eaaimd}{$E_a^{Arr}$}
\newcommand{\eaneb}{$E_a^{NEB}$}

\begin{document}

\title{Combined Experimental and Computational Analysis of Lithium Diffusion in Isostructural Pair \ce{VNb9O25} and \ce{VTa9O25}}

\maketitle

\author{Manish Kumar}
\author{Md Abdullah Al Muhit}
\author{CJ Sturgill}
\author{Nima Karimitari}
\author{John T. Barber}
\author{Hunter Tisdale}

\author{Morgan Stefik*}
\author{Hans-Conrad zur Loye*}
\author{Christopher Sutton*}

\begin{affiliations}
M. Kumar, M. Abdullah, C. Sturgill, N. Karimitari, J.T. Barber, H. Tisdale, M. Stefik, H.-C. zur Loye, C. Sutton\\
Department of Chemistry and Biochemistry, University of South Carolina, South Carolina 29208, United States\\
E-mail: morgan@stefikgroup.com; zurloye@mailbox.sc.edu; cs113@mailbox.sc.edu
\end{affiliations}

\keywords{lithium-ion batteries, Wadsley–Roth phase, lithiation capacity, lithium diffusion, activation barrier}

\begin{abstract}
The increasing demand for fast charging batteries has motivated the search for materials with improved transport characteristics. Wadsley-Roth crystal structures are an attractive class of materials for batteries because lithium diffusion is facilitated by the ReO$_3$-like block structure with electron transport enabled by edge-sharing along shear planes. However, clear structure-property relationships remain limited, making it challenging to develop improved materials with this class of promising compounds. Here, the first lithiation of \ce{VTa9O25} is reported, enabling a direct isostructural comparison with the better-known \ce{VNb9O25}. These materials have similar unit cell volumes and atomic radii yet exhibit different voltage windows, C-rate dependent capacities, and transport metrics. Time-dependent overpotential analysis revealed ionic diffusion as the primary bottleneck to high rate-performance in both cases, however, the corresponding lithium diffusivity for \ce{VNb9O25} was an order of magnitude faster than that for \ce{VTa9O25}. 
These experimental trends aligned well with density functional theory calculations combined with molecular dynamics that show a factor of six faster diffusion in \ce{VNb9O25} as compared with \ce{VTa9O25}.
Nudged elastic band calculations of the probable hopping pathways indicate that \ce{VNb9O25} consistently exhibits a lower activation barrier for lithium diffusion as compared to \ce{VTa9O25}.
Bader charge analysis reveals a larger net charge on Li in \ce{VNb9O25} compared to \ce{VTa9O25} which was attributed to the higher electronegativity of Nb which stabilizes the transition state and lowers the activation energy for diffusion. This stabilization arises from the stronger Coulombic interaction between Li and its coordinated oxygen environment. These materials behave similarly upon lithiation wherein the $a$ and $b$ lattice vectors (which corresponds to the block plane) increase until about 50\% Li concentration and then decreases. However, the electronic structure differs, indicating that \ce{VNb9O25} undergoes a insulator to metal transition at a lower state of charge compared with \ce{VTa9O25}. Overall, this work establishes the role of the cation (Nb or Ta) on the electronic and transport properties during lithiation. 
 
\end{abstract}

\section{Introduction}
With the growing demand for energy storage across a wide range of devices, lithium-ion batteries must offer versatility for fast-charging applications while ensuring long-term durability throughout their lifespan.\cite{frith2023non} To usher in new advancements with battery materials, fundamental studies into structure-property relationships are required for a better understanding of how to design and enhance these materials. The Wadsley-Roth (WR) class of metal oxides have been examined as anode materials and tend to excel in aspects of rate capability, durability through extended cycling, and volumetric energy density.\cite{sturgill2024tailored,griffith2018niobium,zheng2023fast,yang2021wadsley,liu2023recent} For example, \ce{VNb9O25} has been demonstrated to exhibit high-rate capability, often reaching (dis)charge capacities around $\sim$142-156 {\mahg} at a current density of 1 Ag$^{-1}$.\cite{liang2021micro,qian2017high} \ce{VNb9O25} has a theoretical capacity of 945 {\macm} when assuming 1 Li per transition metal (TM), making this WR material comparable to common graphite ($\sim$841 {\macm}). Additionally, \ce{VNb9O25} can maintain 85.6\% of its capacity through 8,000 cycles at $\sim$2.5C (1 Ag$^{-1}$).\cite{liang2021micro} Key reasons for this optimal rate performance and cyclability stem from the WR block structure, which enables rapid diffusion down the block of material, along with shear planes that constrict volumetric expansion during (de)lithiation and preserve the material's crystal structure.\cite{griffith2018niobium,koccer2019cation,griffith2019ionic}  Moreover, calculations show that for \ce{PNb9O25} and \ce{Nb2O5}, among others, the edge-sharing planes of the crystallographic shear structures have shorter metal-metal interatomic spacings with significant orbital overlap that delocalizes electrons and forms electronic conduction pathways that run parallel to the block structure.\cite{griffith2020titanium,yang2021wadsley,griffith2018niobium} 
Moreover, prior work on isostructural analogs primarily focused on comparing \ce{PNb9O25} and \ce{VNb9O25}, and found lower performance in batteries prepared using \ce{VNb9O25},\cite{patoux2002reversible,preefer2020multielectron} underscoring the critical role of composition in WR-phase structures for electrochemical performance and the lithium insertion mechanism. This combination of attributes makes isostructural compounds of \ce{Nb2O5} (e.g, \ce{PNb9O25} and \ce{VNb9O25}) interesting candidates for further studies into their structure-property relationships.\cite{yu2024single}

 In our work, we investigated \ce{VNb9O25} and \ce{VTa9O25}, two compounds that are isostructural with the previously reported \ce{PNb9O25}, to gain an understanding of the effect of substituting the redox-active transition metal, comparing \ce{Nb} versus \ce{Ta}, to elucidate how changes in the primary redox center influence the material's electrochemical behavior. Previous isostructural studies compared the $T[4\times3]$ block structure of \ce{Ta12MoO33} and \ce{Nb12MoO33} where  faster diffusion was found for the tantalate which had slightly smaller diffusion channels that better stabilized transition states during Li-hopping.\cite{muhit2024comparison} 

In this study, an isostructural pair of the $T[3\times3]$ WR phase are compared consisting of \ce{VNb9O25} and \ce{VTa9O25} where again the tantalate had slightly smaller unit cell volume. Batteries were prepared for each material and rate-dependent capacities, sources of overpotential, and transport metrics were compared using lithium half-cells. To compare with experimental measurements, Nudged elastic band (NEB) calculations are used to determine the activation barrier ($E_a$) of different Li hops, which has been used previously to determine the activation barrier between different Li hops in various WR niobates such as \ce{Nb12WO33}, \ce{Nb14W3O44}, \ce{Nb12MoO33}, and \ce{Ta12MoO33}.\cite{koccer2020lithium,muhit2024comparison}
Because of little differences in the structures of \ce{VNb9O25} and \ce{VTa9O25}, a consistent picture for which sites have the most likely Li hops are observed, with the lowest barriers and thus the fastest diffusion along the channel of the block and the slowest diffusion (highest barriers) across shear planes for moving from one block to another.
However, $E_a$ is on average smaller by 25 meV for \ce{VNb9O25} compared with \ce{VTa9O25}. We also performed ab initio molecular dynamics (AIMD) to calculate $D$ directly, from which $E_a$ can be determined which indicated a factor $\sim$ 6 faster $D$ for \ce{VNb9O25}. For \ce{VNb9O25} we modeled the sequence of lithium site filling to identify specific transport pathways active throughout the (de)lithiation process and applied first-principles calculations to provide a detailed insights into redox mechanisms underlying these structural changes. Thus, this work adds information to parse the connections of transport properties in \ce{VNb9O25} and \ce{VTa9O25} with both experimental and computational insights. 

\section{Results}
\subsection{Crystal Synthesis}
\begin{figure}[h!]
    \centering
    \includegraphics[width=0.8\linewidth]{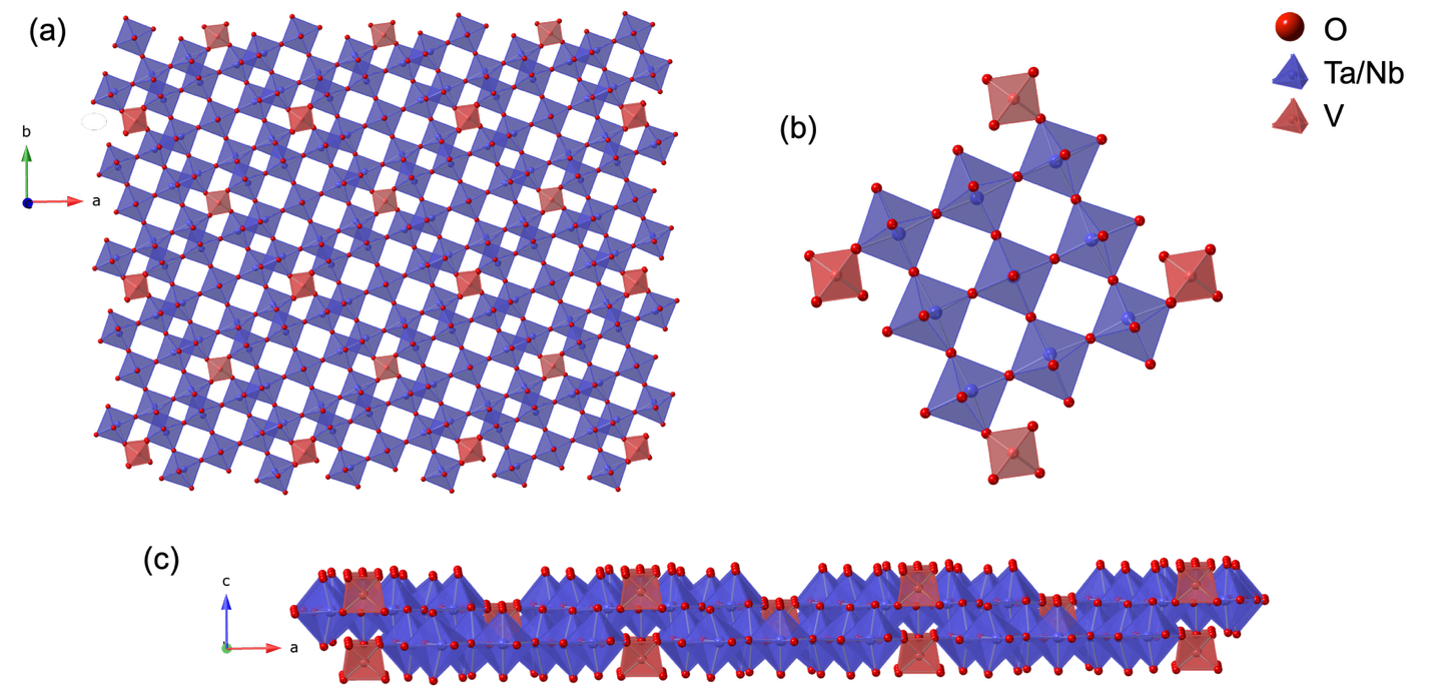}
    \caption{(a) (3$\times$3) blocks are visible from c direction where the blocks are connected via edge sharing in their boundary and also have tetrahedra at the corner. (b) Individual (3$\times$3) block. (c) Blocks lying at different c levels can be observed.}
    \label{fig:cs}
\end{figure}
\ce{VNb9O25} and \ce{VTa9O25} were synthesized by a traditional solid-state route. Since \ce{V2O5} is known to sublimate at high temperatures, the reactions were conducted using pressed pellets in evacuated and sealed fused silica tubes to prevent material loss. Both compounds crystallize in the tetragonal crystal system with the space group $I\bar{4}$, where the asymmetric unit of both compounds contains 4 metal sites, 3 octahedral (2a, 8g, 8g) and one tetrahedral (2c). The $T[3\times3]$ crystal structure in both cases is composed of (3$\times$3$\times\infty$) blocks connected to the neighboring blocks via edge sharing connectivity (Figure~\ref{fig:cs}a). The $3\times3$ blocks are composed of \ce{NbO6} and/or \ce{TaO6} octahedra connected by corner sharing, reflecting the \ce{ReO3} structural motif (Figure~\ref{fig:cs}b). The blocks are offset to each other, forming tetrahedral voids at the junction of the blocks that are occupied by \ce{VO4} tetrahedra (Figure~\ref{fig:cs}c). 

\begin{figure}[h!]
    \centering
    \includegraphics[width=0.8\linewidth]{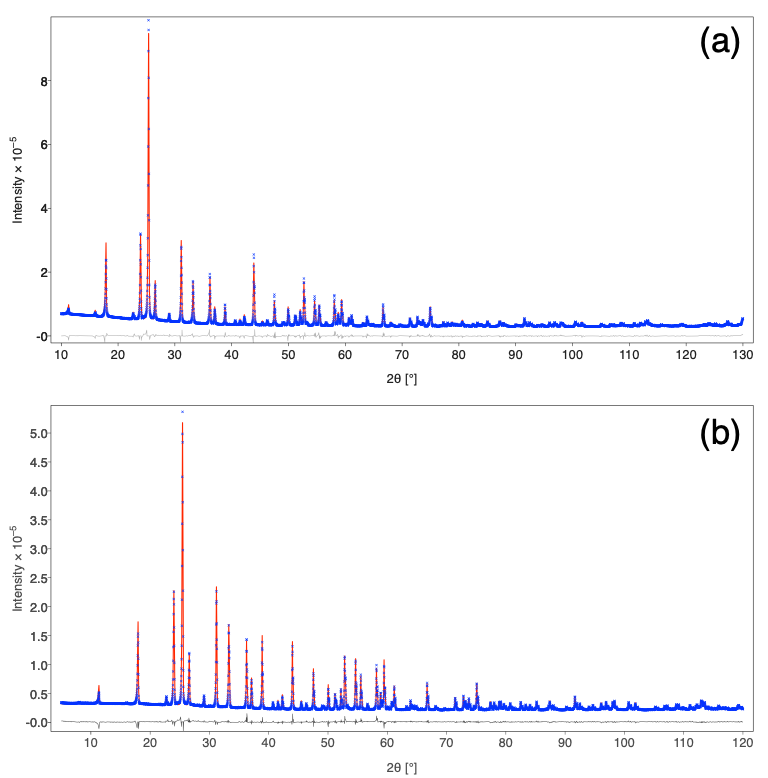}
    \caption{Rietveld refinement of powder X-ray diffraction (XRD) data for (a) \ce{VNb9O25} (Rwp =6.38\%) and (b) \ce{VTa9O25} (Rwp = 3.58\%). The measured intensities are shown as blue crosses, the calculated pattern is shown as a red line, and the residual intensities are shown as a black line. 
    }
    \label{fig:xrd}
\end{figure}

Rietveld refinements were performed to assess sample purity and detect possible anti-site mixing between V/Nb and V/Ta. The final refinements resulted in Rwp values of 6.38\% and 3.58\% for \ce{VNb9O25} and \ce{VTa9O25}, respectively. The Rietveld plots are shown in Figure~\ref{fig:xrd} and the crystallographic information from the Rietveld refinement are listed Table S1. In addition, the atomic coordinates are presented in Table S2. No anti-site mixing was observed and both samples were phase pure \ce{VNb9O25} and \ce{VTa9O25}.

The stability of these materials is estimated using the decomposition enthalpy determined from the convex hull construction, $\Delta H_{d}$, with density functional theory (DFT) calculated formation energies (see Methods for more details). The $\Delta H_{d}$ for \ce{VNb9O25} and \ce{VTa9O25} are -0.38 meV/atom and 0.51 meV/atom.
Further we investigated the stability of anti-site mixed structures (wherein V can occupy the octahedral positions and Nb can occupy the tetrahedral positions) and find that those are higher in energy for both compositions by $\sim$14 and $\sim$21 meV/atom, respectively, for \ce{VNb9O25} and \ce{VTa9O25}. The DFT-optimized lowest-energy structures (i.e., without anti-site mixing) show XRD patterns that compare well with the experimental data (see Figure~\ref{fig:xrd}), particularly in peak intensities. Minor shifts in peak positions are attributed to discrepancies in lattice parameters, with DFT slightly overestimating the unit cell volumes by 5.8\% for \ce{VNb9O25} and 3.7\% for \ce{VTa9O25}.

\subsection{Electrochemical Properties}
\begin{figure}[h!]
    \centering
    \includegraphics[width=0.8\linewidth]{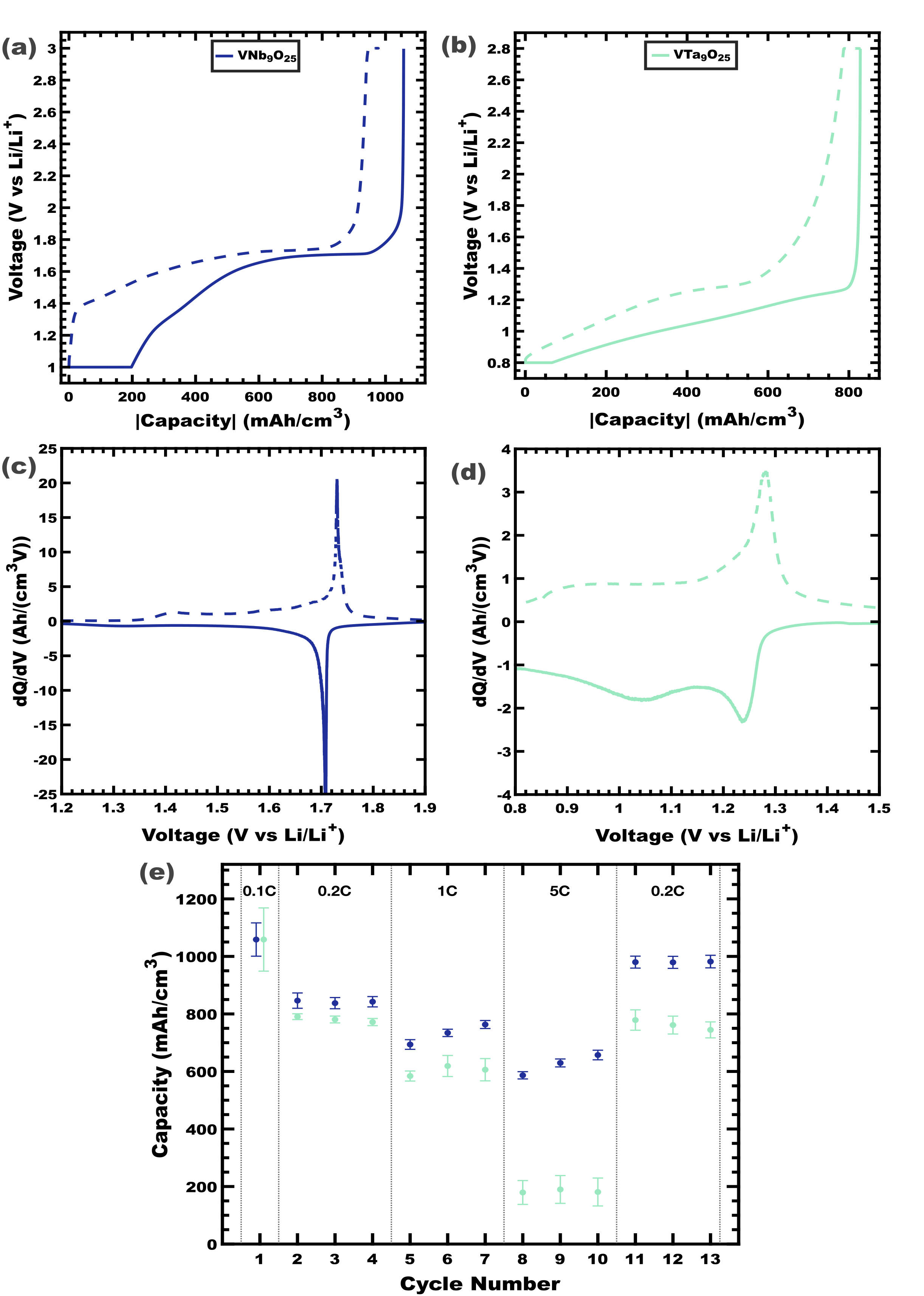}
    \caption{Galvanostatic cycling was performed on coin cells of \ce{VNb9O25} and \ce{VTa9O25}. The (a,b) capacity vs voltage plots are shown at a rate of 0.2C (lithiation, solid line; delithiation, dashed line), along with the corresponding derivative (c,d) {\dqdv} plots. The mean volumetric capacity of \ce{VNb9O25} and \ce{VTa9O25} is shown with (e) varying C-rates. Multiple batteries were measured for each condition, where the error bars correspond to the standard-error-of-the-mean.}
    \label{fig:batteries}
\end{figure}
Coin cells of both \ce{VNb9O25} and \ce{VTa9O25} were assembled vs. lithium to undergo three electrochemical techniques: Galvanostatic charge/discharge (GCD) for rate-dependent capacities, intermittent current interruption (ICI) techniques to examine overpotentials up to 0.2C and separately ICI at 0.1C to measure diffusivity.\cite{chien2023rapid} In previous GCD studies on related Nb/Ta compound pairs, a voltage range of 1.0-3.0 V vs \refli was suitable for the niobate, whereas a shifted redox-window of 0.8-2.8 V vs \refli was needed for the tantalate to reach similar lithiation capacities ($\sim$1 Li/TM) reversibly.\cite{muhit2024comparison} The voltage-capacity profiles of \ce{VNb9O25} and \ce{VTa9O25} were similar with the usual 3 regions found for WR materials, including a (pseudo-)plateau at intermediate voltages (Figure~\ref{fig:batteries}a,b).
The origin of this plateau has been debated, with some previous work suggesting it indicates biphasic behavior.\cite{yang2021wadsley,griffith2017structural,le2024downsizing,liang2024carbon} However, recent studies have shown that solid-solution behavior persists across all lithium contents. Later, entropic potential measurements of several WR phases suggested that this plateau arises from the balance between negative contributions from configurational entropic potential and positive contributions from electronic entropic potential, which are approximately equal in magnitude.\cite{zuras2025investigating} 
The differential capacities (\dqdv) from these galvanostatic datasets reveal further similarities and  differences (Figure~\ref{fig:batteries}c). Both materials exhibit fairly mirror-like {\dqdv} profiles with minor hysteresis between lithiation and delithiation. \ce{VNb9O25} stores a significant amount of charge from 1.6-1.8V vs. \refli, owing to the $Nb^{5+/4+}$ redox couple. A shorter broad peak from 1.2-1.6 V vs. {\refli} likely corresponds to the $Nb^{4+/3+}$ redox couples (Figure~\ref{fig:batteries}c). In contrast \ce{VTa9O25} exhibited broader charge storage peaks in {\dqdv} plots  with the $Ta^{5+/4+}$ redox couple apparent at 1.1-1.3V vs {\refli} (Figure~\ref{fig:batteries}d). The broad peaks for charge storage from  0.8-1.1 V vs {\refli} were ascribed to the $Ta^{4+/3+}$ redox couple. The vandium redox couples of $V^{5+/4+}$ and $V^{4+/3+}$ have been previously shown to irreversibly reduce in the first cycle, with very minimal contribution to charge transfer in the following cycles. Evidence of this can be seen (Figure S1) in the first cycle of both compounds in this study. The galvanostatic lithiation capacities at a rate 0.1C were similar at $\sim$1059 {\macm} for both \ce{VNb9O25} and \ce{VTa9O25} (Figure~\ref{fig:batteries}e). Here, the C-rate corresponds to a current density that is defined inversely with the number of hours needed to reach the theoretical capacity (5C corresponds to 12 min). Progressively higher current densities were examined, where \ce{VTa9O25} generally exhibited more capacity fade than \ce{VNb9O25}. For example, at 1C there was 19.2\% difference (with respect to mean) with 731 {\macm} for \ce{VNb9O25} and 603 {\macm} for \ce{VTa9O25}. Increasing the C-rate to 5C led to 109\% difference with a capacity of 624 {\macm} for \ce{VNb9O25} and 183 {\macm} from \ce{VTa9O25}. Rate-dependent capacity losses are ascribed to overpotentials that limit the amount of current passable at a given current density as constrained by the defined voltage window. The sources of these overpotentials are next discussed.

\subsection{Overpotential Analysis of Rate Limiting Step(s)}
\begin{figure}[H]
    \centering
    \includegraphics[width=0.8\linewidth]{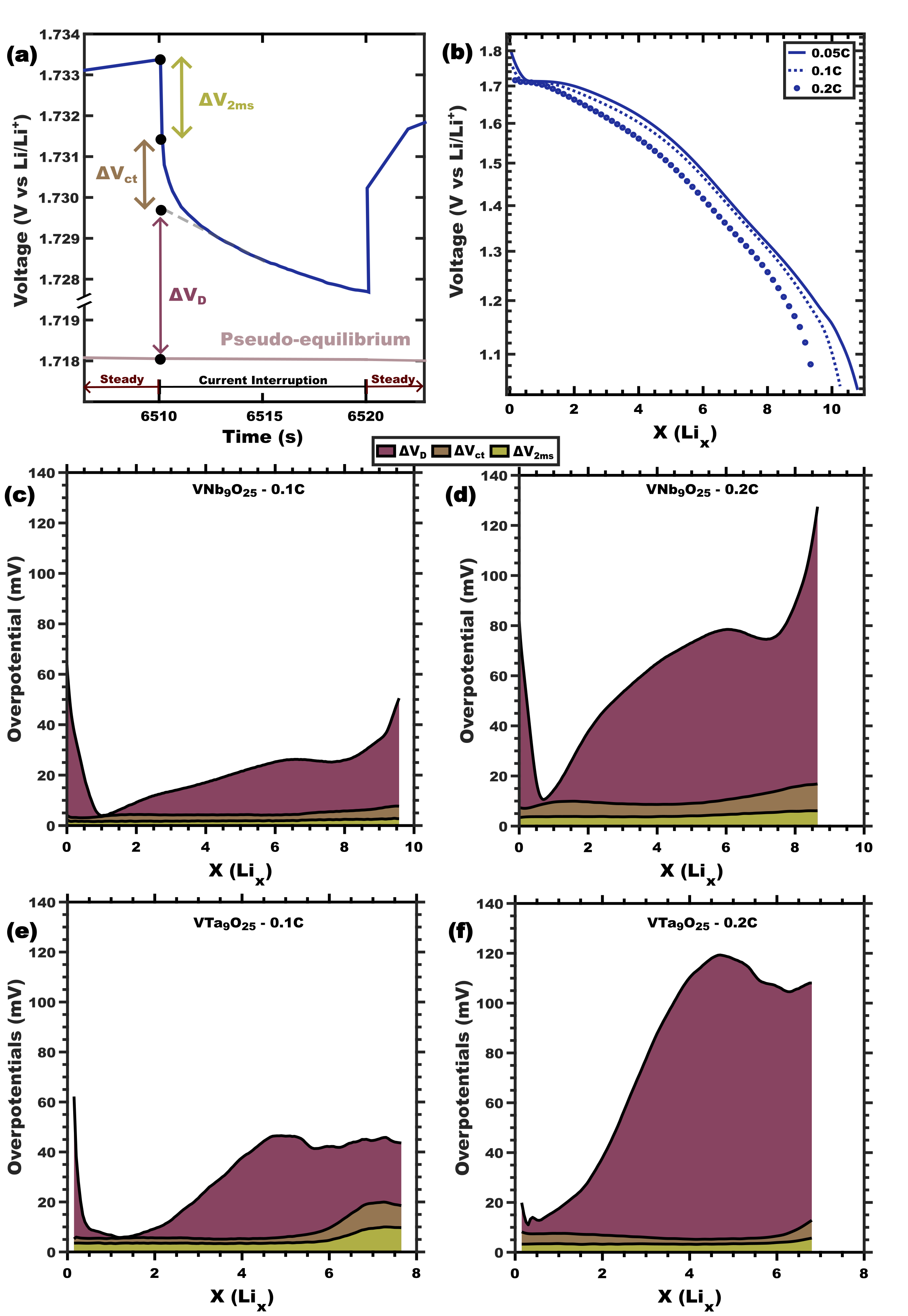}
    \caption{The voltage relaxation during current interruption were parsed into discrete overpotentials based on their time-dependence. (a) Example voltage relaxation profile for \ce{VNb9O25} showing the identified overpotentials: \vms, \vct, and {\vd}. The magnitude of {\vd} required comparison to a pseudo-equilibrium reference (turquoise line) measured at 0.05C with current interuptions to derive the iR-correction (compared at equivalent \lix). (b) Example voltage-capacity plots for \ce{VNb9O25} at different C-rates. The overpotentials for \ce{VNb9O25} and \ce{VTa9O25} are compared at rates of 0.1C (c) and 0.2C (d) across the voltage window.}
    \label{fig:overpotentials}
\end{figure}

An ICI approach was used to investigate the sources of overpotential in each material by examining the time-dependent voltage relaxations. ICI approaches are usually used to derive transport metrics such as 
ionic diffusivity, cell resistance, and diffusion resistance where small current densities corresponding to $\leq$ 0.1C are important to avoid electrolyte concentration gradients such that overall diffusion resistance is dominated by solid-state diffusion alone.\cite{chien2023rapid} 
The voltage relaxation profiles from ICI experiments have been used to parse different processes based on their time-dependent responses~\cite{geng2021intermittent} where example data from \ce{VNb9O25} is shown in Figure~\ref{fig:overpotentials}a. The instantaneous voltage relaxation is often measurable within 2 ms ($\Delta V_{2ms}$) and is attributed to the sum of 1) electronic resistance of the electrodes, 2) electronic resistance of the current collectors, and 3) bulk ionic resistance of the electrolyte (not ionic diffusion). On longer time scales, often 1 to 5 s, the voltage change from ionic diffusion is fitted as a response that is linear with $t^{0.5}$ (see gray dashed line in Figure~\ref{fig:overpotentials}a) to derive diffusion resistance and diffusivity. Extrapolating this $V \propto t^{0.5}$ fit back to $t=0$ (relative to current interruption) corresponds to the sum of $\Delta V_{2ms}$ and {\vct} for charge transfer (this sum was previously termed $\Delta V_{reg}$~\cite{geng2021intermittent}). Finally, the magnitude of {\vd} requires reference to the equilibrium potential at the given state of lithiation. {\vd} was measured by comparison with a  pseudo-equilibrium (Figure~\ref{fig:overpotentials}a purple line) cycle with a current density corresponding to 0.05C with periodic current interruptions to determine the iR-corrected pseudo open circuit potential~\cite{chien2023rapid}(pseudo-equilibrium V compared at the same extent of lithiation \lix). The iR-corrected pseudo open circuit potential was noted to have hysteresis between lithiation and delithiation (roughly 40-80 mV for \ce{VNb9O25} and 70-150 mV for \ce{VTa9O25}) thus the presented overpotentials are underestimated. In this way ICI voltage relaxation profiles are interpretable in terms of fundamental processes. Furthermore, there is no apparent issue with applying this approach at higher current densities with recognition that diffusion contributions may not be purely from solid-state diffusion. This approach was used to analyze the sources of overpotential for \ce{VNb9O25} and \ce{VTa9O25} at current densities corresponding to 0.1C and 0.2C. The resulting overpotential profiles are shown as stacked plots such that the sum of all overpotentials is also apparent (Figure~\ref{fig:overpotentials}c-f). For the 0.2C examples of both samples, the diffusion overpotentials {\vd} were generally dominant with similarly scaled and minor contributions from {\vms} and {\vct}. Also, for a given extent of lithiation, the overpotentials were generally larger for \ce{VTa9O25}. Clear trends with increasing \vd were found when comparing 0.1C to a higher current density of 0.2C (Figure~\ref{fig:overpotentials}e-f). It is apparent that both \ce{VNb9O25} and \ce{VTa9O25} have diffusion constraints as the dominant source of overpotential. Given the dominance of {\vd} overpotential losses, the material diffusivities were next compared in detail.

\subsection{Diffusivity Analysis}
\begin{figure}[h!]
    \centering
    \includegraphics[width=1.0\linewidth]{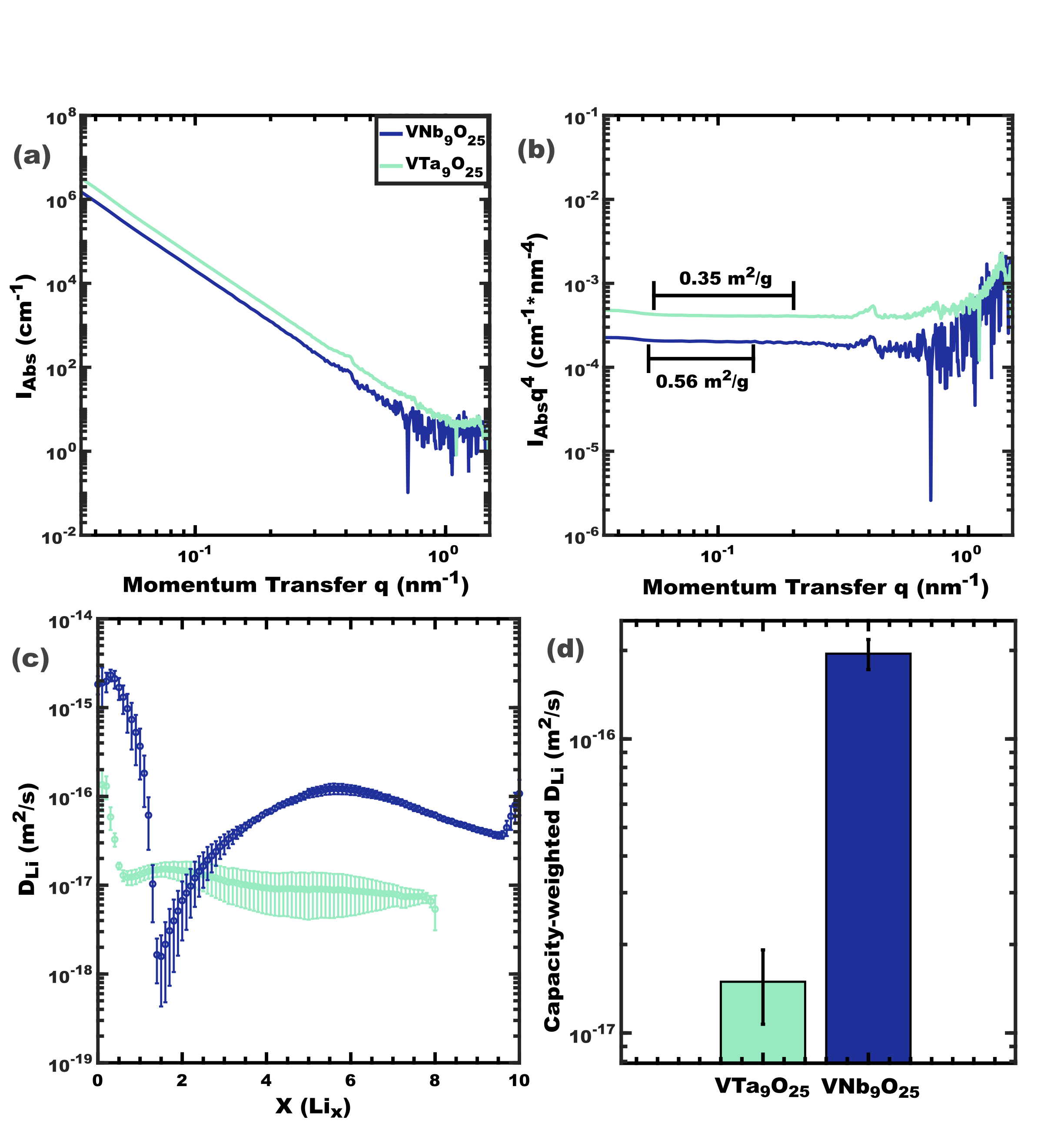}
    \caption{(a) Absolute intensity SAXS data where (b) analyzed with Porod plots to determine the mass specific surface areas from the Porod region of constant {$Iq^4$} value (indicated in black). (c) Diffusivities of \ce{VNb9O25} and \ce{VTa9O25} were calculated across a range of lithiation extents using the ICI technique where error bars correspond to the sample-to-sample variation. (d) Capacity-weighted diffusivities  were calculated as figures-of-merit that take into account the full collection of measured values across triplicated measurements (error bars correspond to error-of-the-mean).}
    \label{fig:expdiff}
\end{figure}

Given the dominance of the {\vd} overpotential, the samples were further compared in terms of the diffusivity values. The standard ICI technique at 0.1C was used to determine the diffusion resistance of each 
sample across the accessible ranges of lithiations. Converting diffusion resistances into the more generalizable diffusion coefficients requires accounting for the intercalation length scale, which is typically derived from measurements of the mass-specific surface area. As recently discussed in detail,\cite{sturgill2025capacity} the percent error of the surface area measurement is doubled in the calculation of diffusivity where error minimization is crucial to determine accurate values. Ensemble methods such as BET (Ar-based) or SAXS Porod analysis (effective thickness corrected) are suitable to minimize error in reported values. Here, the latter approach was used where absolute intensity SAXS data were analyzed using Porod plots (Figure~\ref{fig:expdiff}a,b) to determine the mass specific surfaces areas of 0.561 $\pm$ 0.008  m$^2$g$^{-1}$ for \ce{VNb9O25} and 0.352 $\pm$ 0.003 m$^2$g$^{-1}$ for \ce{VTa9O25}. With these values, the diffusivities were calculated for each sample across the lithiation range where \ce{VNb9O25} generally had higher diffusivity values than \ce{VTa9O25}, except for {\lix} with $1<x<2$ (Figure~\ref{fig:expdiff}a). Capacity-weighted transport metrics are important when comparing such materials that undergo second-order phase transitions upon lithiation since all of the $D(x)$ values contribute to the overall (dis)charging process.\cite{sturgill2025capacity} Furthermore cases such as this comparison with a cross-over of values makes a figure-of-merit necessary for a fair comparison. Thus, the capacity-weighted diffusivity values were calculated for comparison with $D_{avg} =$ $1.95 \times 10^{-16}$ {\mss} for \ce{VNb9O25} and $D_{avg} =$ $1.49 \times 10^{-17}$ {\mss} for \ce{VTa9O25} (Figure~\ref{fig:expdiff}c). Thus, \ce{VNb9O25} had a factor of 13.1 higher capacity-weighted diffusivity than \ce{VTa9O25}. These substantially different diffusivities explain the different rate-dependent capacities found with GCD experiments (Figure~\ref{fig:batteries}d). 

\subsection{Computational Analysis}
\begin{figure}[h!]
    \centering
    \includegraphics[width=1.0\linewidth]{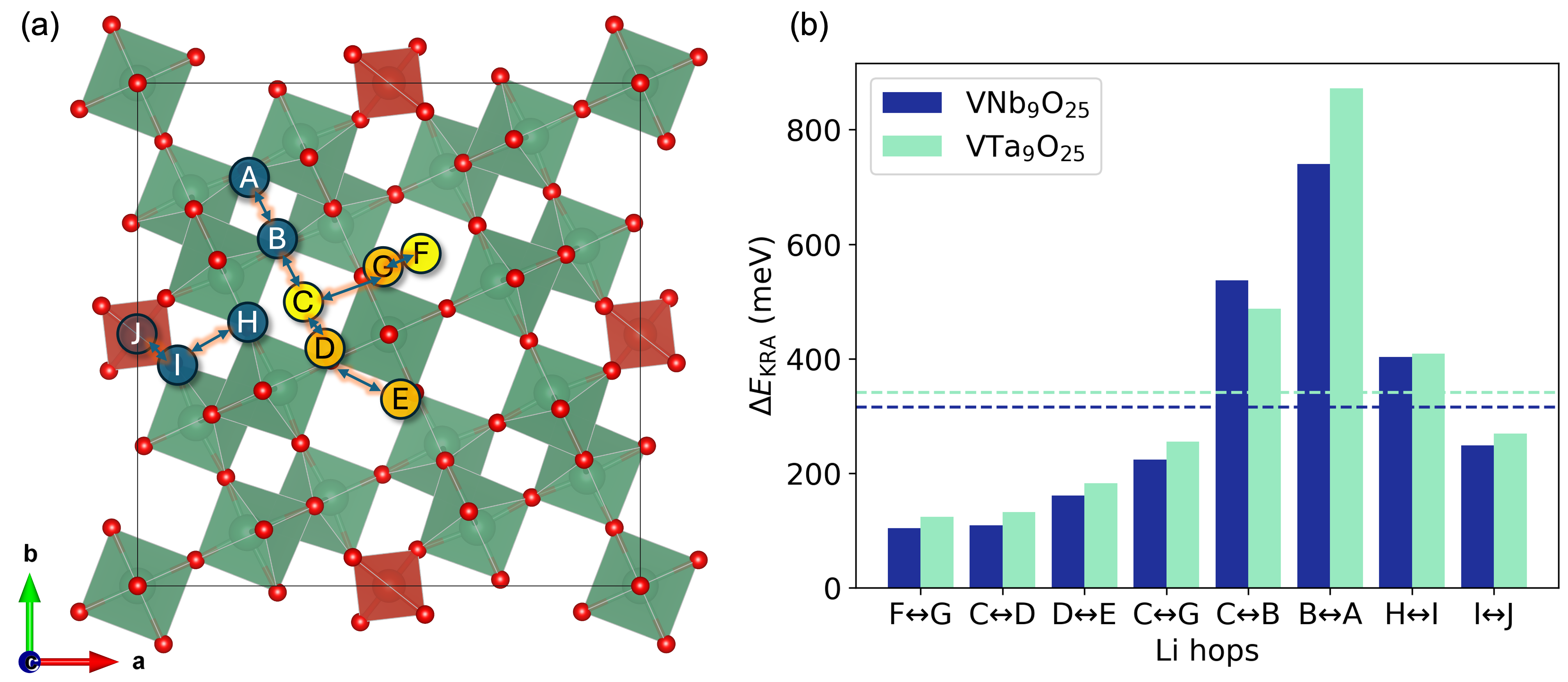}
    \caption{(a) Li sites and hops between them. The ``pocket" sites (A,B,H,I) are 5-fold coordinated, lying at the periphery of the block along the shear planes. The other 4-fold coordinated ``window" sites (C,D,E,F,G) found inside the block are described as ``horizontal" (C,F) or ``vertical" (D,E,G) depending on the orientation of the window relative to the plane of the block. The ``cavity" site (J) is present in-between the \ce{VO4} tetrahedra. (b) {\eaneb} for lithium ion motion. The colored dashed lines show the means of the activation barriers.}
    \label{fig:sites}
\end{figure}

To compare with experiment, we calculated the Li diffusion coefficients ($D$) in both materials, \ce{VNb9O25} \ce{VTa9O25}, using AIMD simulations at a fixed a lithium concentration of 50\% ($x=5$) with respect to transition metals. Because of well-known issues of trapping and thus undersampling states with wide-ranging activation energies, $D(x=0.5)$ at $T=300~K$ were estimated from Arrhenius fits to AIMD-calculated values over the temperature range of $T = 500–900~K$, as shown in Figure S2 (see Methods section for a detailed description of the approach). The Arrhenius plots for \ce{VNb9O25} and \ce{VTa9O25} are included in the figure, and the corresponding activation energies ({\eaaimd}) derived from these fits are listed in Table~\ref{tab:diff_ea}. The results indicate that \ce{VNb9O25} exhibits a diffusion coefficient approximately 6 times higher than that of \ce{VTa9O25}, consistent with experimental observations.
\begin{table}[h!]
    \centering
    \begin{tabular}{c|c|c}
    \hline
    Compound & $D$ (m$^2$s$^{-1}$) & {\eaaimd} (meV)\\
    \hline
      \ce{VNb9O25}   & $3.65\times10^{-11}$ & 164\\
      \ce{VTa9O25}   & $6.39\times10^{-12}$ & 226\\
      \hline
    \end{tabular}
    \caption{Diffusion coefficient, $D$, obtained from AIMD simulations, and activation barrier, {\eaaimd}.}
    \label{tab:diff_ea}
\end{table}
To understand the reasons for this faster diffusion in \ce{VNb9O25}, we determined the individual activation barriers of different Li hops in both \ce{VNb9O25} and \ce{VTa9O25} using the nudged elastic band (NEB) method,\cite{henkelman2000climbing} which has been applied in several previous works to examine Li diffusion in WR phases.\cite{muhit2024comparison,koccer2020lithium}
For different Li hops, we identified six symmetrically distinct lithium sites. The six nonequivalent sites, A, C, D, H, I, and J are shown in Figure~\ref{fig:sites}a. The 5-fold coordinated ``pocket" sites (A,B,H) lie at the periphery of the block along the shear planes, while pocket site I lies near \ce{VO4} tetrahedra. The other ``window" sites (C,D,E,F,G) found inside the block are 4-fold coordinated and described as ``horizontal" (C,F) or ``vertical" (D,E,G) based on the orientation of the window relative to the plane of the block. The ``cavity" site (J) is occupying the site between \ce{VO4} tetrahedra. We adopted the nomenclature of different sites as reported by Ko{\c{c}}er et al.\cite{koccer2020lithium} These sites have a multiplicity of four except for the J site, which has a multiplicity of one per formula unit.
The hops between different Li-sites in \ce{VNb9O25} and \ce{VTa9O25} are shown in Figure~\ref{fig:sites}a.
To remove the directional dependence of the activation barrier, {\eaneb}, between initial and final states, we calculate it as: 
\begin{equation}
    E_a^{NEB} = E_\mathrm{TS} - \frac{1}{2}(E_\mathrm{i}+E_\mathrm{f})
\end{equation}
where $E_\mathrm{TS}$ is the energy of the transition state or saddle point obtained from the energy profile, $E_\mathrm{i}$ and $E_\mathrm{f}$ are the energies of initial and final states.
{\eaneb} for different hops are given in Table~\ref{tab:my_label}; Overall, these calculations show a lower activation barrier for Li diffusion in \ce{VNb9O25} by on average 25 meV compared to \ce{VTa9O25} (Figure~\ref{fig:sites}b), consistent with the AIMD calculated {\eaaimd} derived from the Arrhenius fit.  
The smallest barriers of {\eaneb} 104 and 124 meV from NEB are predicted for Li hops from horizontal-to-vertical window sites (F$\xleftrightarrow{}$G) in \ce{VNb9O25} and \ce{VTa9O25}, respectively. These types of hops (F$\xleftrightarrow{}$G and C$\xleftrightarrow{}$D) facilitate long-range diffusion along the channel.\cite{koccer2020lithium} Within the $ab$ plane, the Li can jump within the same block with next small barriers of 161 and 183 meV (D$\xleftrightarrow{}$E) in \ce{VNb9O25} and \ce{VTa9O25}, respectively. Whereas, Li diffusion from C$\xleftrightarrow{}$B (i.e., from window to the pocket site) has an {\eaneb} of 537 and 487 meV in \ce{VNb9O25} and \ce{VTa9O25}, respectively, and thus should be a slower process compared with Li hopping within the channel. The largest barriers ({\eaneb}  = 740 and 872 meV in \ce{VNb9O25} and \ce{VTa9O25}, respectively) occur for Li diffusion across the shear plane from pocket-to-pocket Li hops (B$\xleftrightarrow{}$A), i.e., from one block to another block. These results indicate that Li diffusion is expected to occur fastest within the channel due to a lower {\eaneb}, leading to a higher diffusion coefficient ($D$) in \ce{VNb9O25} compared to \ce{VTa9O25}. However, at low Li concentration, the Li from pocket site could jump into the channel due to the overpotential, thereby it could also lead to faster diffusion at small concentration. Overall, our results for {\eaneb} are similar in magnitude to other WR shear structures;\cite{koccer2020lithium,muhit2024comparison} however, we find a consistently smaller {\eaneb} for \ce{VNb9O25} in comparison to \ce{VTa9O25} for all Li hops examined, but C$\xleftrightarrow{}$B, which is the opposite case to our previous work comparing a T[4x3] Nb/Ta pair.\cite{muhit2024comparison}

Although there is qualitative agreement between our simulations and the experimental results, we note that the calculations differ from the experimental setup in at least two key aspects. First, in the simulations, Li hopping occurs under a constant diffusion overpotential during (de)lithiation ($\nabla \phi(x)$), which is often not the case experimentally (Figure~\ref{fig:overpotentials}). Second, Li diffusion in reality varies with Li concentration, whereas in both the NEB and AIMD calculations, we assume a fixed concentration due to computational constraints. More specifically, for AIMD simulations were performed at a single Li concentration, $x=0.5$. Connecting these {\eaneb} values for individual sites to the experimental measured $D(x)$ trends in Figure \ref{fig:overpotentials}a, which varies significantly across the range of lithiations, requires knowledge of the Li filling sequence. Any Li ordering that may result has the potential to alter the allowed sequence of diffusion paths, which can change the resulting diffusion rates which we will discuss now. 

\begin{table}[h!]
    \centering
    \begin{tabular}{c|c|c}
        \hline
        Hops & \ce{VNb9O25}& \ce{VTa9O25} \\
        \hline
        F$\xleftrightarrow{}$G & 104 & 124 \\
        C$\xleftrightarrow{}$D & 109 & 132 \\
        D$\xleftrightarrow{}$E & 161 & 183 \\
        C$\xleftrightarrow{}$G & 224 & 255 \\
        C$\xleftrightarrow{}$B & 537 & 487 \\
        B$\xleftrightarrow{}$A & 740 & 872 \\
        H$\xleftrightarrow{}$I & 403 & 409 \\
        I$\xleftrightarrow{}$J & 249 & 269 \\
        \hline
    \end{tabular}
    \caption{Li hops activation barrier, {{\eaneb}} (in meV) between different sites labeled in Fig \ref{fig:sites}.}
    \label{tab:my_label}
\end{table}

\begin{figure}
    \centering
    \includegraphics[width=0.5\linewidth]{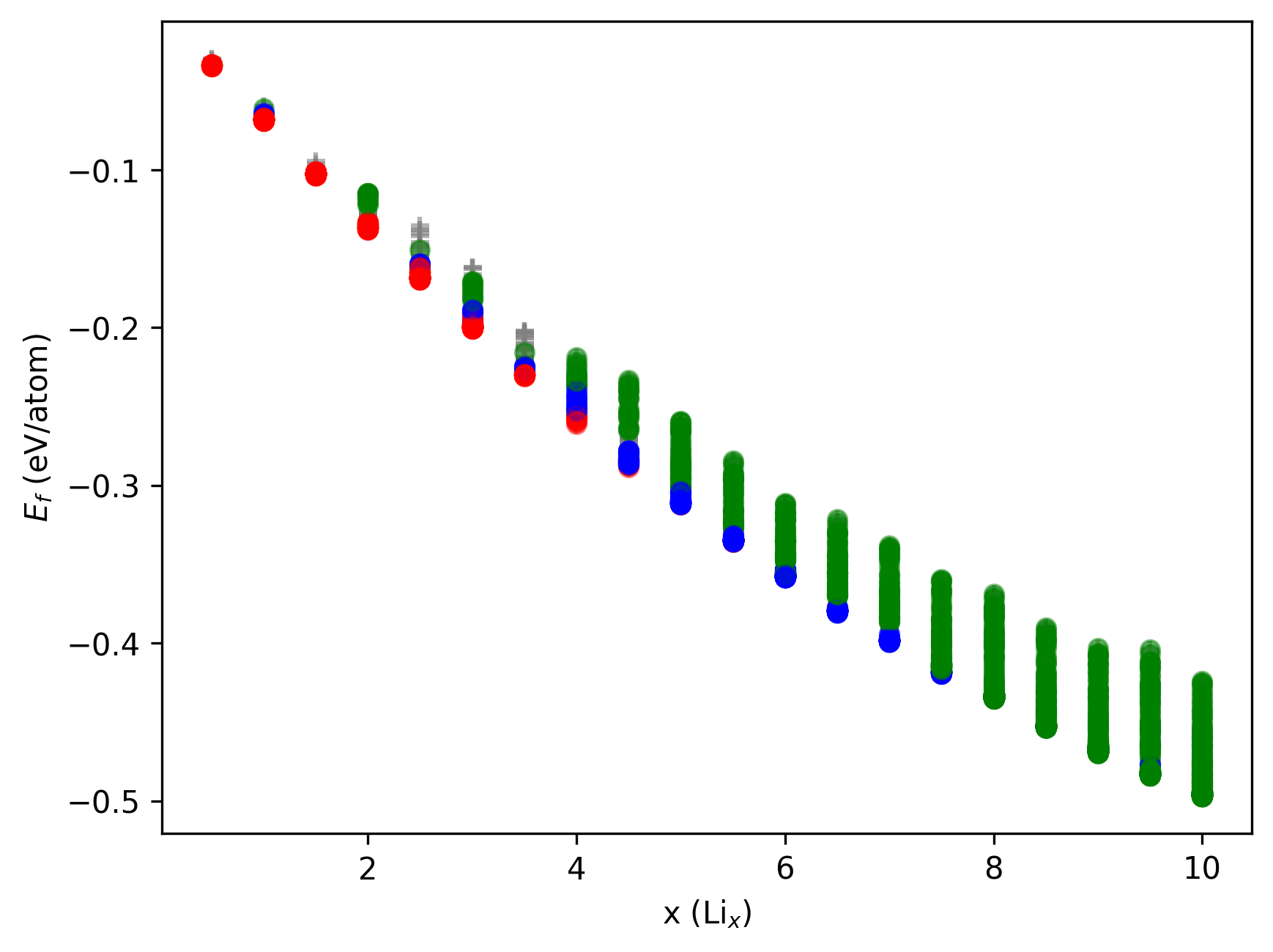}
    \caption{Formation energy of 7142 Li orderings in \ce{Li$_x$VNb9O25} calculated using CE+MC. Here, grey color shows all the orderings, while red color highlights the Li occupation only at pocket sites, near shear plane (B/A) and near the tetrahedra (I), blue color shows the Li occupation at another pocket site near the corner (H) and the cavity site (J) in addition to Li occupation at pocket sites (B/A/I) and green color indicates Li occupation at all the six sites (pocket, window and cavity sites).}
    \label{fig:ce-mc}
\end{figure}

The sequence of Li filling (or Li configurations) is explored using cluster expansion (CE) trained to the MLIP optimized structures, which is combined with the Metropolis Monte Carlo (MC) simulations at $T=300$ K (see Methods). 
The CE+MC approach determines the low energy Li-configurations across different $x$ values, which is shown in Figure~\ref{fig:ce-mc}. 
Our results indicate that at lower concentration ($x \leq 4$), low-energy configurations states contain the Li at pocket sites B/A near the shear plane and I-sites near the tetrahedra, while high-energy configurations partially occupy all the sites.
At $x=4$, half of the I-sites and half of the B/A sites are occupied by Li in the lowest-energy configuration. These are followed by the occupation of the cavity site J and pocket site H. The horizontal (C/F) and vertical window sites (D/E/G) start to be occupied in low energy configurations at $x\sim6$. This sequence of site filling at different Li concentrations is similar to what was calculated by Saber et al.\cite{saber2021role} for \ce{PNb9O25} using CE. They had also seen that at lower Li concentrations, pocket sites are occupied, and at higher Li concentrations, window sites are getting occupied in low-energy configurations. Further, we find that the distribution of Li among all sites become more probable with increase in Li concentration, increasing the disorder of Li with concentration (see SI Figure S3). 

For \ce{VNb9O25}, the experimental diffusivity trends correlate to this calculated sequence of changes in the lowest-energy lithium configuration:Ordering of the lithium configurations preferentially places lithium ions in specific sites which constrains the available types of hops for diffusion due to different {\eaneb} values. 
To examine the dynamics of Li starting from distinct sites, we performed MD simulation using the MLIP for a single Li in \ce{VNb9O25} ($x=0.17$) at $T=300$ K. Due to a large thermal activation energy of these hops and neglecting any effect of the external applied potential, Li starting in pocket or cavity sites is trapped near the block edge, whereas it can travel across the channel when starting from windows sites until trapped later in a pocket/cavity site. This can be seen from the trajectory lines of Li, shown in Figure S4. These results of the Li dynamics from MD are consistent with the higher {\eaneb} for the jumps from the pocket/cavity sites and lower {\eaneb} of jumps from the window site determined (discussed above). 

From these perspectives, energetically well-separated sites (non-degenerate) are anticipated to have the diffusivity depend on the probability of site occupancy as well as the probability of neighboring site being vacant which leads to a parabolic $D(x)$ trend that is concave-down. This expectation matches the experimental data for \ce{VNb9O25} $D(0<x<1.5)$ which corresponds to preferential filling of pocket sites. Notably, it is unexpected that the pocket site occupation would lead to the most rapid diffusion observed experimentally since those sites have the highest activation energies for diffusion (Figure~\ref{fig:expdiff}). However, the high overpotential for lithiating these sites (Figure ~\ref{fig:overpotentials}c,d), could be responsible for activating transport from these sites. Galvanostatic conditions maintain constant current density by shifting the applied voltage, as constrained by the set voltage window. The first lithiums added to \ce{VNb9O25} (~2 V vs \refli) are far from the cutoff voltage (1.0 V vs. \refli), where the battery cycler has sufficient voltage remaining for overpotential to drive the motion of such "trapped" states. Notably, the pronounced decrease at $D(x\sim 1.5)$ is consistent with a crowding effect (insufficient fraction of vacant sites) combined with an inability to energetically access other diffusion paths. The subsequent $D(1.5<x<9.5)$ trends for \ce{VNb9O25} appear like a convolution of several such parabolas, consistent with a lower degree of partitioning for higher extents of lithiation as the various window sites become progressively occupied and then crowded. Notably the overall experimental trends for \ce{VTa9O25} are quite similar (Figure ~\ref{fig:expdiff}c) in character. From these perspectives, it is remarkable that the fastest diffusion observed experimentally is associated with the pocket sites that have the highest activation energies for Li hopping.

The {\eaneb} in \ce{VNb9O25} results in a higher $D$, but a key question remains: why is {\eaneb} lower in \ce{VNb9O25} compared to \ce{VTa9O25}? Notably, these materials are isostructural and exhibit very similar crystal structures, yet their calculated $D$ values differ significantly.
To investigate the origin of the lower activation barrier in \ce{VNb9O25}, we calculated the effective charge of Li in the three different positions: initial, final, and transition states (TS) of the hopping process. Across all 8 paths, the charge of Li is larger in \ce{VNb9O25} than \ce{VTa9O25}, which will stabilize the TS for \ce{VNb9O25} more than \ce{VNTa9O25} because of stronger Coulombic attraction between Li and the surrounding O atoms, leading to a lower activation barrier observed in \ce{VNb9O25}. Although Nb and Ta have similar ionic radii, the difference in charge may be due to Nb having slightly higher electronegativity. 
\begin{table}[]
    \centering
    \begin{tabular}{ccccccc}
        \hline
    \multirow{2}{*}{Hops} &
      \multicolumn{3}{c}{\ce{VNb9O25}} &
      \multicolumn{3}{c}{\ce{VTa9O25}}  \\
        & Initial & TS & Final & Initial & TS & Final\\
        \hline
        F$\xleftrightarrow{}$G & 0.38 & 0.67 &0.64 & 0.40&  0.59&0.53\\
        C$\xleftrightarrow{}$D & 0.40 & 0.64 &0.63 & 0.67& 0.60 &0.61\\
        D$\xleftrightarrow{}$E & 0.62 & 0.59 &0.63 & 0.61&  0.56&0.61\\
        C$\xleftrightarrow{}$G & 0.40 & 0.61 &0.64 & 0.67&  0.46&0.53\\
        C$\xleftrightarrow{}$B & 0.40 & 0.63 &0.43 & 0.67& 0.52 &0.35\\
        B$\xleftrightarrow{}$A & 0.43 & 0.44 &0.44 & 0.35&  0.37&0.37\\
        H$\xleftrightarrow{}$I & 0.46 & 0.51 &0.45 & 0.46& 0.46 &0.39\\
        I$\xleftrightarrow{}$J & 0.45 & 0.60 &0.59 & 0.39& 0.56 &0.61\\ 
        \hline
    \end{tabular}
    \caption{Bader charges $(\abs{e})$ on Li atom for initial, transition and final state during the hops.}
    \label{tab:bader}
\end{table}

\begin{figure}
    \centering
    \includegraphics[width=1.0\linewidth]{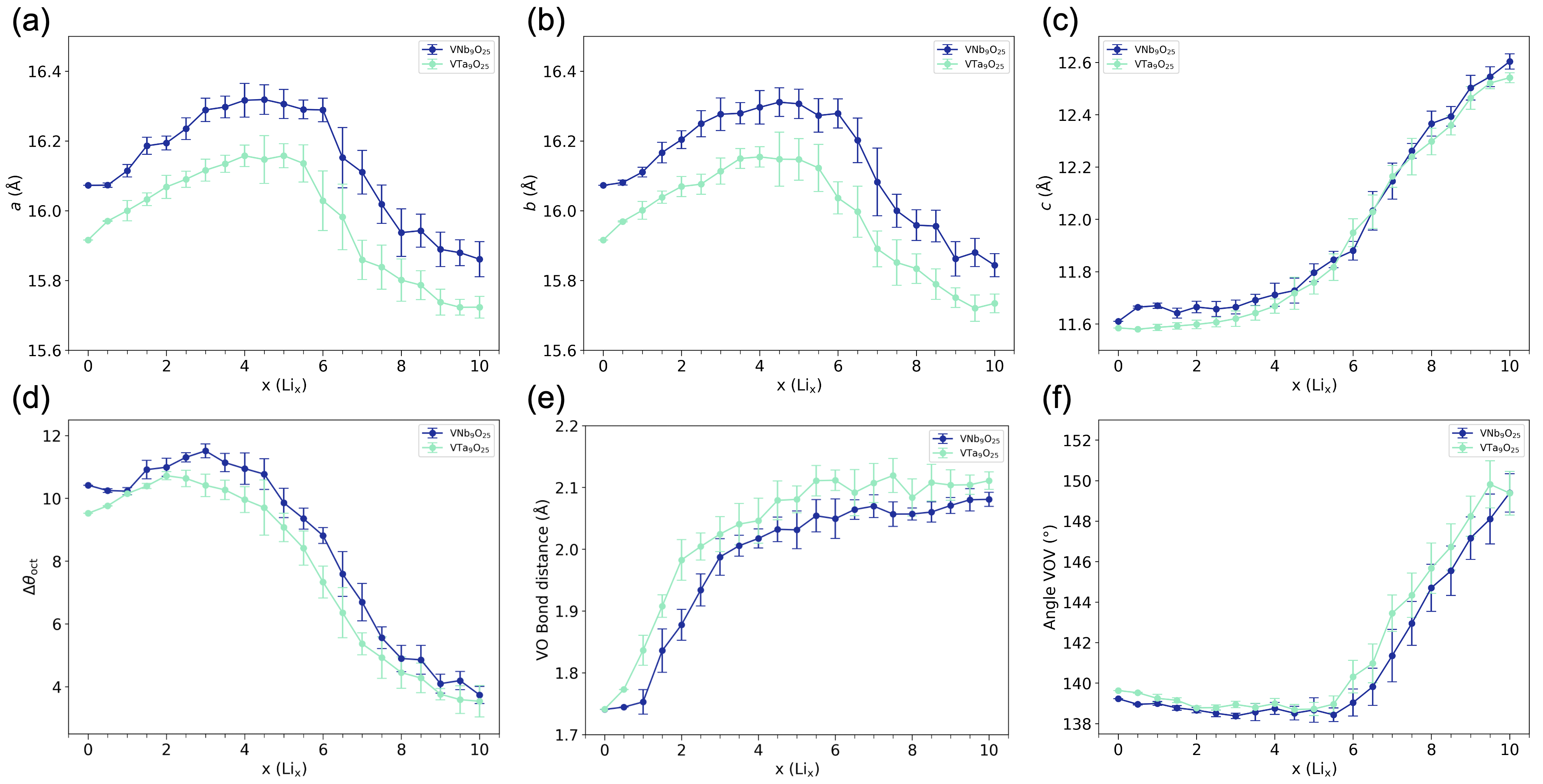}
    \caption{Comparison of structural evolution of average lattice parameters, octahedral distortion ($\Delta \theta_{oct}$), VO bond length ($d_{V-O}$), and VOV angle as a function of Li concentrations in \ce{VNb9O25} and \ce{VTa9O25}.}
    \label{fig:struc}
\end{figure}

In addition to the change in the relative energy of the Li configurational states upon increasing Li, both the crystal structure and electronic structure are also affected. As has been discussed extensively before for WR structures,\cite{qian2017high,muhit2024comparison} the lattice parameters first increase in the $a$ and $b$ directions up to approximately $x \sim 0.5$, after which they begin to decrease (see Figure ~\ref{fig:struc}). In contrast, the lattice parameter increases continuously in the $c$ direction (along the channel) with increasing $x$. Similar to what has been pointed out before, this decrease in the $a$ and $b$ directions is due to an increased symmetry in the cation octahedra based on the average $\Delta \theta_{oct}$ (Figure~\ref{fig:struc}), which is the bond angle O-Nb/Ta-O variance divided by the square of the mean of the bond angles.\cite{koccer2019cation}
As can be seen in Figure~\ref{fig:struc}f, the increase in VOV angles follows monotonically with the increase in the $c$ lattice parameter. This trend is consistently observed for \ce{VNb9O25} and \ce{VTa9O25}; however, \ce{VNb9O25} remains consistently larger than that of \ce{VTa9O25}. Moreover, for iso-structural \ce{PNb9O25} and \ce{Nb10O25}, Saber et al.\cite{saber2024redox,saber2021role} have previously identified a decrease in the Nb-Nb distances along the shear-plane with increasing lithium concentrations, which was attributed to the formation of Nb-Nb dimers. 
We determined the distribution of Nb-Nb/Ta-Ta distances as a function of Li concentration (see SI Figure S5) for lowest 10 orderings. Consistent with these previous studies, a decrease on average, in the bond distances between the Nb octadehdral at the edge-sharing sites (see Figure S5a). Conversely, the Nb-Nb distances at the corner-sharing sites increase with increasing Li concentration (see Figure S5b). However, we also observe that the V-O bond distance ($d_{V-O}$) increases with Li concentration with a dramatic jump from  $d_{V-O} \sim$ 1.7 {\AA} to 2.0 {\AA} (with larger $d_{V-O}$ is observed overall for \ce{VNb9O25} compared with \ce{VTa9O25}), which has not been discussed in the literature previously.

These structural changes have a significant impact on the electronic structure, as indicated in the changes in the electronic partial density of states (pDOS) of lithiated structures from $x=0.5$ to $10$. The pDOS of structures without Li (i.e., $x = 0$) show that Nb/Ta-$d$ orbitals are contributing to conduction band minimum (CBM) and O-$p$ states are contributing to valence band maximum (VBM), having a band gap energy of $\sim$ 2 and 2.6 eV, for \ce{VNb9O25} and \ce{VTa9O25}, respectively (see Figure S6). As Li ions intercalate, electrons fill the conduction bands.  At a very low concentration of Li ($x=0.5$), \ce{VNb9O25} becomes metallic as the Fermi level shifts to CBM. \ce{VTa9O25} remains a semiconductor until the Li content reaches $x=2.5$, at which point it undergoes a transition to a metal. Another qualitatively different trend in the pDOS between these two compounds is the appearance of states formed from hybridization of V-$d$ and O-$p$ in the VB at low Li concentration ($x=0.5$) in the case of \ce{VTa9O25}, but only appear at $x\ge1.5$ in \ce{VNb9O25}. Moreover, \ce{VNb9O25} exhibits a greater number of available electronic states compared to \ce{VTa9O25} beyond $x=2.5$, as indicated by the higher intensity peak in \ce{VNb9O25} at conduction band.

\section{Conclusion}
In summary, we have systematically performed the combined experimental and computational study to compare the lithiation properties in isostructural \ce{VNb9O25} and \ce{VTa9O25}. Despite having similar crystal structure, these exhibits different electrochemical and transport properties. 
The lithiation capacity is found to be higher in \ce{VNb9O25} as compared to \ce{VTa9O25} at most of the C-rates ($\geq 0.2$C).
The diffusion of Li ions in \ce{VNb9O25} is found to be an order magnitude faster than that of \ce{VTa9O25}. This difference aligns well with the lower activation barrier of Li hops and higher diffusion coefficient of Li in \ce{VNb9O25} calculated using NEB and AIMD. Moreover, the transition state of the Li hops is more stabilized in the case of \ce{VNb9O25}, leading to lower activation barrier. 
The diffusivity trend across different Li concentrations is well explained with lithium ordering and activation barrier of different hops. Moreover, the structure evolution with lithiation corroborates with the corresponding changes in electronic structures. We observe that \ce{VNb9O25} becomes metallic at dilute Li concentration, while that transition happens at $x=2.5$ in \ce{VTa9O25}. Overall, these results are expected to guide the rational design of high-power lithium ion batteries using Wadsley-Roth crystallographic shear structures.

\section{Experimental}
\threesubsection{Chemicals}
\ce{V2O5} (Alfa Aesar, 99.6\%), \ce{Nb2O5} (Alfa Aesar, 99.9985\%), \ce{Ta2O5} (Alfa Aesar, 99\%), Super P Conductive Carbon Black (“carbon black,” MSE), N-Methyl-2-pyrrolidone (NMP, Sigma-Aldrich, 99.5\%), Poly-vinylidene fluoride (PVDF, MTI, 99.5\%), Acetone (Fisher, 99.5\%), Lithium hexafluorophosphate in 1:1 ethylene carbonate/dimethyl carbonate (Sigma-Aldrich), Lithium metal (MSE, 99.9\%) were used as received.

\threesubsection{Synthesis of Polycrystalline Powders of \ce{VNb9O25} and \ce{VTa9O25}}
Powders of \ce{Nb2O5} and \ce{Ta2O5} were mixed with \ce{V2O5} in the appropriate stoichiometric ratios, ground using mortar and pestles, and pressed into a pellet using 2 tons of pressure. The pellets were sealed in evacuated ($\sim 10^{-4}$ torr) fused silica tubes using a methane-oxygen torch. The samples were heated to 1000 $\degree$C and 1200 $\degree$C at 600 $\degree$C/hr, dwelled for 24 hours, at which point the furnace was turned off. The pellets were removed from the tubes and ground into powders for phase purity characterization and property measurements. The white color of the samples indicated that all the metals were in their highest oxidations states. 

\threesubsection{Powder X-ray Diffraction}
Powder X-ray diffraction (PXRD) data was collected on the polycrystalline samples to confirm phase purity. PXRD data was collected on a Bruker D2 PHASER diffractometer using Cu K$\alpha$ radiation ($\lambda$ = 1.5418 \AA) over the 2$\mathrm{\theta}$ range $5-65\degree$ with a step size of 0.02$\degree$.

\threesubsection{Rietveld Refinement of Powder Samples}
PXRD data were collected on polycrystalline samples of \ce{VNb9O25} and \ce{VTa9O25}  on a Rigaku Ultima IV diffractometer utilizing a sealed tube X-ray source (Cu K$\alpha$) operated in the Bragg-Brentano geometry. Data were collected over a 2$\mathrm{\theta}$ range of $5-120\degree$ for \ce{VTa9O25} and $10-130\degree$ for \ce{VNb9O25} at a step size of 0.02$\degree$. To confirm the phase identities and purities of both samples, Rietveld refinements were performed on both patterns using Bruker TOPAS version 6 operated via jEdit 5.5.0.  Initial models were taken from previously reported structure models in ICSD (space group: I$\bar4$). For \ce{VTa9O25}, the zero point, Chebychev background parameters, unit cell parameters, Thompson-Cox-Hastings profile parameters, axial divergence, and scale factor were refined. The temperature factors and atomic positions were fixed according to the initial models. For \ce{VNb9O25}, the background was fit manually and the sample displacement, unit cell parameters, Thompson-Cox-Hastings profile parameters, axial divergence, and scale factor were refined. The temperature factors for the one V and all 3 Nb sites were set as equal and refined as one variable; the temperature for all 7 O sites were also set to equal and refined as a single variable. Significant preferred orientation was present in both patterns and were accounted for by refining high-order spherical harmonics.

\threesubsection{Electrochemistry}
Slurry-based electrodes were prepared for lithium half-cell measurements in a coin cell. Each slurry was prepared by thoroughly grinding an 8:1 ratio of active material to carbon black in a mortar and pestle by hand. Next, a 25 gml$^{-1}$ solution of PVDF in NMP was added to the oxide/carbon mixture to create an overall ratio of 8:1:1 of oxide, carbon black, and PVDF. The resulting NMP-based mixture was stirred with a magnetic stir bar for two hours to eliminate agglomerates and ensure thorough mixing. Copper foils were cleaned by wiping using Kimwipes moistened with acetone. After that, the slurry was doctor bladed onto the copper foils using a blade height of 15 $\mu$m. The wet films were immediately dried on a hot plate set to 85 $^{\circ}$C for at least 15 mins. Then, the films were further dried in a heated vacuum oven set to 100 $^{\circ}$C for 21 hrs. Dried electrodes were then punched into 12 mm discs, weighed in triplicate, and brought into an argon glovebox for assembly into coin cells. The 2032-type coin cells were assembled using 1.0 M \ce{LiPF6} in ethylene carbonate/dimethyl carbonate electrolyte, Whatman glass fiber separators, and 16 mm lithium metal chips, 2×0.5 mm stainless steel spacers, and 1 stainless steel wave spring. Galvanostatic cycling and intermittent current interruption (ICI) techniques~\cite{chien2023rapid} were performed using the Biologic BCS-810. The C-rate was defined based on one lithium per transition metal, corresponding to a 1C current density of 208 {\mahg} and 129 {\mahg} for \ce{VNb9O25} and \ce{VTa9O25} respectively. For rate capability testing, the \ce{VNb9O25} cells were cyclically (de)lithiated between 1.0-3.0 V at constant current (C-rate capacity), likewise, the \ce{VTa9O25} cells were cyclically (de)lithiated between 0.8-2.8 V at constant current (C-rate capacity). Each (de)lithiation was ended with a voltage hold section to allow equilibration before the next step. At the end of this C-rate testing, the ICI technique was used to calculate diffusivity using the following equation:
\begin{equation}
D=\frac{4}{\pi}\left(\frac{V}{A}\frac{\frac{\Delta E_{OC}}{\Delta t_{I}}}{\frac{dE}{d\sqrt{t}}}\right)^2
\end{equation}
where $V$ is the molar volume of the electrode material, $A$ is the mass specfic surface area, $E_{OC}$ is the open circuit potential, $\Delta t_I$ is the period of constant current applied between OCP measurements, $E$ is the potential of the electrode, and $t$ is the step time. The ICI technique was used during galvanostatic cycling at a rate of 0.1C. During ICI, each 300 s galvanostatic period was ended with an interruption by changing the current to zero for 10 s. The transient voltage was monitored during this interruption to derive the lithium diffusivity using the linear fit in root time of the 1-5 s portion for the slope. The overpotential drop from 0-2ms following the current pause was recorded using a modulobat technique to monitor voltage change immediately following the pause. The charge transfer overpotential was determined relative to \vms based on extrapolation of the fit diffusion trend back to t=0.\cite{chien2023rapid,geng2021intermittent} The \vd was determined by comparison to a pseudoequilibrium GCD measured at 0.05C by aligning the voltages with each extent of lithiation (\lix). Overpotentials were gathered as detailed above that occurred during the transient portion of ICI measured using C-rates of 0.1C and 0.2C.\cite{geng2021intermittent,stolz2024practical,chen2020development} Custom MATLAB code was used to analyze all ecltrochemical data. Note that each condition was measured in triplicate to present average battery values along with the error-of-the-mean. The specific surface areas used in diffusivity calculations were derived from SAXS Porod analysis.

\threesubsection{Small-Angle X-ray Scattering}
Small-Angle X-ray Scattering (SAXS) measurements were completed for Porod surface area analysis on both samples of \ce{VNb9O25} and \ce{VTa9O25}. The measurements were performed on a SAXSLab Ganesha at the South Carolina SAXS Collaborative (SCSC). A Xenocs GeniX 3D microfocus source was used with a copper target to generate a monochromatic X-ray beam with a wavelength of 0.154 nm and a flux of 5.54E7 counts per second. A Pilatus detector captured both the scattered signal and also measured the directly transmitted beam intensity for standardless absolute scattering. Kapton tape was used to mount samples and all measurements included subtraction of the Kapton background signal using the SAXSGUI software. SAXS data are presented in terms of the magnitude of the momentum transfer vector q where $q=4\pi sin(\theta)/\lambda$ where $2\theta$ is the total scattering angle. All SAXS measurements were carried out using the absolute scattering intensity using an approach elegantly elaborated by Spalla to calculate the apparent thickness of sample. This approach is superios to caliper measurements and accounts for incomplete powder compactness.\cite{spalla2003analysis,lu2022specific} The X-ray absorption coefficients of 474 $cm^{-1}$ and 961 $cm^{-1}$ at 8.04 keV for \ce{VNb9O25} and \ce{VTa9O25} respectively (NIST calculator~\cite{nist_calc2}) were combined with transmission measurements to calculate the apparent sample thickness used in absolute scattering calculations. Porod plots were analyzed to calculate specific surface area starting from the surface to volume ratio defined by:
\begin{equation}
\Sigma=\frac{\lim_{q\to\infty}I_{abs} q^4}{2 \pi(\Delta \mathrm{SLD})^2}
\end{equation}
Where SLD is the scattering length density of the material and the limit of $q \to \infty$ of $I_{abs}q^4$ is a constant-value segment of the Porod region. The X-ray scattering length density contrast of 3.55E-5 $A^{-2}$ and 5.00E-5 $A^{-2}$ was determined using the NIST calculator for \ce{VNb9O25} and \ce{VTa9O25} respectively.\cite{nist_calc1} The plateau value of $Iq^4$ in the Porod region was calculated as an average value with the q-range extended until the intensity error was less than 1\%. The sample specific surface area was then calculated using:
\begin{equation}
 S=\frac{\Sigma}{\rho} 
\end{equation}
where $\rho$ is the sample bulk density (4.54 {\gcm} for \ce{VNb9O25} and 7.37 {\gcm} for \ce{VTa9O25}).

\threesubsection{DFT Methods} \label{DFT_Methods}
Spin-polarized DFT calculations were performed using the Vienna Ab initio simulation package (VASP).\cite{kresse1996efficiency,kresse1999ultrasoft} The projector-augmented wave (PAW) pseudopotentials~\cite{kresse1999ultrasoft,blochl1994projector} were used to describe the interactions between electrons and ions for all of the species. The specific PAW pseudopotentials used for Li (Li\_sv), Nb (Nb\_pv), Ta (Ta\_pv), V (V\_pv), and O (O) are consistent with pymatgen's~\cite{ong2013python} MPRelaxSet.  A generalized gradient approximation, specifically, the Perdew-Burke-Ernzerhof (PBE)~\cite{perdew1996generalized} exchange-correlation functional with Hubbard parameter $U$ was used for the calculations. $U$ of 3.25 used for V was chosen from the pymatgen python package’s MPRelaxSet, which tabulates $U$ parameters calibrated using the approach described by Wang et al.\cite{wang2006oxidation} The planewave kinetic energy cutoff of 520 eV was used for basis set expansion. For Brillouin zone integration, a $\Gamma$-centered $k$-point mesh with a grid density of at least 1000/(atoms/unit cell) was used. The geometries were optimized until the forces on each atom were smaller than 0.01 eV/\AA\, and the energy converged within the threshold of $1\times10^{-6}$ eV. We optimized \ce{VNb9O25} and \ce{VTa9O25} with DFT+$U$ in the tetragonal I$\bar{4}$ space group (Number 82) with Nb (\ce{VNb9O25}) or Ta (\ce{VNb9O25}) occupying the octahedral sites at the block center and edges and V occupying the tetrahedral sites located at the block corners. 
Moreover, for anti-site mixing of octahedral and tetrahedral sites, we generate the special quasirandom structure (SQS)~\cite{zunger1990special,van2013efficient} where the block centers and the block corners can by occupied by V and Nb (\ce{VNb9O25}) or Ta (\ce{VTa9O25}), respectively.

Nudged elastic band (NEB)~\cite{henkelman2000climbing} calculations were performed in cubicized supercells ($1\times 1 \times 3$) of \ce{VNb9O25} and \ce{VTa9O25}, containing 210 atoms, using the VTST tools.\cite{henkelman2000climbing,henkelman2000improved} A single Li atom was inserted at each unique position shown in Figure~\ref{fig:sites}. In the case of input structures for NEB calculations, only the atomic positions were optimized to keep the cell fixed during the transition state search. 4-6 images were used to resolve the path between different hops. The NEB calculations were stopped once the force on each image was smaller than 0.05 eV/\AA.

Ab initio molecular dynamics (AIMD) simulations were performed in cubicized supercells ($1\times 1 \times 3$) of \ce{VNb9O25} and \ce{VTa9O25}, containing 210 non-Li atoms, and 30 Li atoms (50\% of cations). To obtain the Li insertion, 1000 configurations were generated using a simple algorithm that randomly places Li atoms in the cell and accepted only those structures with Li bond distances $> 1 \,\mathrm{\AA}$. These structures were subsequently optimized using the MACE-MP0.\cite{batatia2023foundation} From this set, the lowest-energy configuration was selected. AIMD simulations were performed across five different temperatures in increments of 100 K (500-900 K), each for 7.6 ps. The simulations used a Langevin thermostat in the $NVT$ ensemble, with a time step of 1 fs. The first 3 ps was kept for equilibration. From the rest 4.6 ps trajectory, dividing it into subsamples of length 0.77 ps, the diffusion coefficient ($D$) was calculated from the Einstein relation:
\begin{eqnarray}\label{eins_rel}
    D &=& \frac{1}{2d}\lim_{t\to\infty} \frac{d}{dt} \frac{1}{N}\sum_{\mathclap{i=1}}^{N} \langle \abs{r_i(t+t_0)-r_i(t_0)}^2\rangle_{t_0}\\
    &=& \frac{1}{2d}\lim_{t\to\infty} \frac{d}{dt} \langle\Delta r^2 (t)\rangle
\end{eqnarray}
where $N$ is the number of diffusing particles, $d$ is the dimensionality of the diffusion, and $r_i(t+t_0)$ represents the position of atom $i$ at time $t$, following an equilibration time $t_0$.
Note that this equation is valid for so-called self-diffusion, which measures how a particle (e.g., Li ion) diffuses through a material in the absence of concentration gradients. The room temperature ($T=300$ K) $D$ values were extracted by extrapolating the Arrhenius equation, which describes how the diffusion rate depends on temperature:
\begin{equation}\label{my_first_eqn}
    D = D_0 e^{\left(\frac{-E_a}{k_BT}\right)}
\end{equation}
The activation energy ($E_a$) can be determined from the slope of the Arrhenius equation. 
To ensure a reasonable computational cost for the AIMD simulations, the planewave kinetic energy cutoff was reduced to 450 eV.

To determine the ground states of \ce{Li$_\mathrm{X}$VNb9O25}, as a function of concentration, we construct a cluster expansion (CE) model of the formation energy, using the Python Package CELL.\cite{rigamonti2024cell} For training dataset, we build supercells of parent lattice containing six unit cells, forming a dataset of 860 structures, having different Li concentrations up to 100\% with respect to transition metals. The formation energy $E_{f,i}$ is calculated using the following equation
\begin{equation}
    E_{f,i} = (E_i-E_\mathrm{host}- x_\mathrm{Li}E_\mathrm{Li})/\mathrm{N}
\end{equation}
where $E_i$ is the energy of host cell containing Li atoms, $E_\mathrm{host}$ is the energy of host cell, and $E_\mathrm{Li}$ is the energy of the bulk Li. 
$x_\mathrm{Li}$ is the number of Li atoms and N is the total number of atoms in the host cell. The energies are obtained using fine-tuned machine learned interatomic potential, MACE.\cite{batatia2023foundation} To obtain the CE model, we consider a set of clusters containing the 1-, 2-, 3-, and 4-point clusters, making a set of 366 clusters. The cross-validated root mean squared error (CV-RMSE) is 5.1 meV/atom. The CV-RMSE is minimum at 96 number of clusters. Using this CE, Metropolis Monte Carlo simulations were performed at $T=300$ K with 1 million steps at different Li concentrations.

\medskip
\textbf{Supporting Information} \par
Supporting Information is available from the Wiley Online Library or from the author.

\medskip
\textbf{Acknowledgements} \par
All authors acknowledge DOE support (DE-SC0023377). This work made use of the South Carolina SAXS Collaborative. This research used computational resources sponsored by the Department of Energy’s Office of Energy Efficiency and Renewable Energy and located at the National Renewable Energy Laboratory. This research also used resources of the National Energy Research Scientific Computing Center (NERSC), a Department of Energy Office of Science User Facility using NERSC award ERCAP0029139.

\medskip
\textbf{Conflict of Interest} \par
The authors declare no conflict of interest.

\medskip
\textbf{Author Contributions} \par
M.K., M.A., and C.J.S. contributed equally to this work.

\medskip
\textbf{Data Availability Statement} \par
The data that support the findings of this study are available from the corresponding author upon reasonable request.

\bibliographystyle{MSP}
\bibliography{main.bib}

\end{document}